\documentclass[10pt]{article}

\usepackage{amsfonts}
\usepackage{amsmath}
\usepackage{authblk}
\usepackage{bm}
\usepackage{color}
\usepackage[d]{esvect}
\usepackage[margin=2cm]{geometry}
\usepackage{graphicx}
\usepackage[colorlinks=true]{hyperref}
\usepackage{subcaption}
\usepackage{xparse}
\usepackage{xr}

\newcommand{\ttl}{Experimental data over quantum mechanics simulations for inferring the repulsive exponent of the Lennard-Jones potential in Molecular Dynamics}
\newcommand{\aut}{
    \author[1]{Lina Kulakova}
    \author[1]{Georgios Arampatzis}
    \author[1,2]{Panagiotis Angelikopoulos}
    \author[1]{Panagiotis Chatzidoukas}
    \author[3]{Costas Papadimitriou}
    \author[1,*]{Petros Koumoutsakos}
    \affil[1]{Computational Science and Engineering Laboratory, Clausiusstrasse 33, ETH Zürich, CH-8092, Switzerland}
    \affil[2]{Department of Mechanical Engineering, University of Thessaly, Pedion Areos, GR-38334 Volos, Greece}
    \affil[3]{Currently at D.E.Shaw Research LLC, 10016 NY USA}
    \affil[*]{petros@ethz.ch}
}

\makeatletter
\newcommand*{\addFileDependency}[1]{
  \typeout{(#1)}
  \@addtofilelist{#1}
  \IfFileExists{#1}{}{\typeout{No file #1.}}
}
\makeatother

\DeclareDocumentCommand \myexternaldocument { m } {
    \externaldocument{#1}
    \addFileDependency{#1.tex}
    \addFileDependency{#1.aux}
}

\newcommand*\smref[1]{Supplementary Material~\ref{#1}}

\newcommand{\beginsupplement}{
    \clearpage
    \setcounter{table}{0}
    \renewcommand{\thetable}{S\arabic{table}}%
    \setcounter{figure}{0}
    \renewcommand{\thefigure}{S\arabic{figure}}%
    \setcounter{equation}{0}
    \renewcommand{\theequation}{S\arabic{equation}}%
    \setcounter{section}{0}
    \renewcommand{\thesection}{S\arabic{section}}%

	\newcommand{\smaketitle}{
		{\LARGE Supplementary Material: \ttl}%
		\vskip10pt
		{\large Lina Kulakova, Georgios Arampatzis, Panagiotis Angelikopoulos, Panagiotis Chatzidoukas, Costas Papadimitriou, and Petros Koumoutsakos}
		\vskip20pt
	}
}

\renewcommand{\vec}{\bm}

\newcommand{\R}{\mathbb{R}}

\newcommand{\N}{\mathcal{N}}
\newcommand{\U}{\mathcal{U}} % uniform
\newcommand{\LN}{\mathcal{L}} % log-normal
\newcommand{\TN}{\mathcal{T}} % truncated normal

\DeclareDocumentCommand \pr { m o } { \IfNoValueTF{#2} {p(#1)}{p(#1 \,|\, #2)} }

\newcommand{\cnt}{\, ,} % continue
\newcommand{\stp}{\, .} % stop

\newcommand{\M}{\mathcal{M}} % model
\newcommand{\MU}{\mathcal{U}} % uniform model
\newcommand{\MLN}{\mathcal{L}} % log-normal model
\newcommand{\MTN}{\mathcal{T}} % truncated normal model

\newcommand{\vx}{\vec{x}}
\newcommand{\y}{d} % one data point
\newcommand{\vy}{\vec{d}} % data
\newcommand{\vvy}{\vv{\vy}} % vector of data
\newcommand{\C}{\vec{\Sigma}} % covariance matrix

\newcommand{\p}{\vartheta} % parameters
\newcommand{\vp}{\vec{\vartheta}} % vector of parameters
\newcommand{\pe}{\varepsilon} % LJ eps
\newcommand{\ps}{\sigma} % LJ sigma
\newcommand{\pp}{p} % LJ exponent
\newcommand{\pn}{\sigma_n} % std of error

\newcommand{\hp}{\psi} % hyper-parameters
\newcommand{\vhp}{\vec{\psi}} % vector of hyper-parameters

\newcommand{\lb}{\left(}
\newcommand{\rb}{\right)}

\myexternaldocument{si} % xr

\begin{document}

%%%%%%%%%%%%%%%%%%%%%%%%%%%%%% FRONT PAGE %%%%%%%%%%%%%%%%%%%%%%%%%%%%%%%%%%%%%%
\title{\ttl}
\aut
\date{\today}
\begin{abstract}
The Lennard-Jones (LJ) potential is a cornerstone of Molecular Dynamics (MD) simulations and among the most widely used computational kernels in science. The potential models atomistic attraction and repulsion with century old prescribed parameters ($q=6, \; p=12$, respectively), originally related by a factor of two for simplicity of calculations. We re-examine the value of the repulsion exponent through data driven uncertainty quantification. We perform  Hierarchical Bayesian inference on MD simulations of argon using experimental data of the radial distribution function (RDF) for a range of thermodynamic conditions, as well as dimer interaction energies from quantum mechanics simulations. The experimental data suggest a repulsion exponent ($p \approx 6.5$), in contrast to the quantum simulations data that support values closer to the original ($p=12$) exponent. Most notably, we find that predictions of RDF, diffusion coefficient and density of argon are more accurate and robust in producing the correct argon phase around its triple point, when using the values inferred from experimental data over those from quantum mechanics simulations. The present results  suggest the need for data driven recalibration of the LJ potential across MD simulations.
\end{abstract}
\maketitle
%%%%%%%%%%%%%%%%%%%%%%%%%%%%%% END FRONT PAGE %%%%%%%%%%%%%%%%%%%%%%%%%%%%%%%%%%

%%%%%%%%%%%%%%%%%%%%%%%%%%%%%% INTRODUCTION %%%%%%%%%%%%%%%%%%%%%%%%%%%%%%%%%%%%
\section{Introduction} \label{sec:intro}
The Lennard-Jones (LJ) potential is one of the centerpieces in Molecular Dynamics (MD) simulations,  the key computational method for studying atomistic phenomena across Chemistry, Physics, Biology and Mechanics. Despite the widespread use of MD simulations, a usually overlooked fact is that the classic LJ potential involves a century old and rather ad-hoc prescribed repulsion exponent. In this study we demonstrate that this parameter needs to be modified in order to enhance the predictive capabilities of MD simulations.

The structure of the LJ potential depends on the inter-atomic distance $(r)$ and consists of two parts: an attractive term $-r^q$ and a repulsive term that models the Van der Waals forces and a repulsive term $r^{p}$ that models the Pauli repulsion.  While the exponent $q=6$ has a theoretical justification \cite{Jones:1924} the $p=12$ exponent has no physical justification and it was chosen for simplicity as it can be computed as the square of the attractive term.
In addition, two scaling parameters $\ps$ and $\pe$ control the shape of the potential.
The $\pe$ and $\ps$ parameters have been the subject of numerous calibration studies~\cite{Barker:1971,Rahman:1964,Rowley:1975,White:1999} and more recently the subject of Bayesian inference techniques~\cite{Cailliez:2011,Angelikopoulos:2012}. Bayesian Uncertainty Quantification (UQ) employs experimental data and provides  a probability distribution of the parameters. The parameter uncertainty can then be propagated by the model in order to obtain robust predictions on a quantity of interest~\cite{Angelikopoulos:2012,Hadjidoukas:2015a}. In cases where the data sets correspond to different inputs for the system, e.g. different thermodynamic conditions, the use of Hierarchical Bayesian (HB) methods provides a  stable method for UQ~\cite{Wu:2015,Wu:2016}. 

Here we employ a HB method to  infer the parameters $(\pe,\ps,\pp)$. In particular, we infer systematically the LJ 6-$\pp$ exponents using experimental data from Radial Distribution Function (RDF) and data from quantum simulations of argon. In the past, several values of the exponent $\pp$ of the LJ 6-$\pp$ potential, ranging from 10 to 20, have been considered~\cite{Galliero:2006}. The authors calibrated  using pressure and viscosity data for various thermodynamic conditions and concluded that the exponent 12 is the best choice.  
Here, we perform HB inference for the LJ 6-$\pp$ parameters of argon based on experimental RDFs of liquid argon and saturated argon vapor for six different temperature and pressure pairs. We present a rigorous model selection process for the LJ $6-12$ vs LJ $6-\pp$ potentials for each of the cases and perform robust posterior predictions for the diffusion coefficient and density. The $6-\pp$ potentials inferred with RDF experimental data are being compared with those inferred using data from quantum simulations. We conclude that the 6-$\pp$ potential inferred with RDF data is the only potential that can simulate a wide variety of thermodynamic conditions. 
Moreover, we find that the most likely values for the exponent are $\pp \approx 6.5 $, strongly differing from the value of $\pp = 12$  that is  being used.  We remark that our results have bveen obtained in the case of a simple system. However we consider that they offer significant evidence that the repulsive exponent should be reconsidered when  the parameters of the LJ potential are being fitted to data. 
%%%%%%%%%%%%%%%%%%%%%%%%%%%%%% END INTRODUCTION %%%%%%%%%%%%%%%%%%%%%%%%%%%%%%%%

\section{Results} \label{sec:uq_res}
%%%%%%%%%%%%%%%%%%%%%%%%%%%%%% RESULTS %%%%%%%%%%%%%%%%%%%%%%%%%%%%%%%%%%%%%%%%%
We first calibrate the parameters of the classical LJ 6-12 potential. This inference is denoted as $B_{12, R}$. Subsequently we include the exponent of the repulsion term into the parameter set (inference $B_{\pp, R}$) and perform model selection for the LJ 6-12 and LJ 6-$\pp$ force fields. Finally, we perform a HB inference for each of the potentials using the methodology from Ref.~\cite{Wu:2016}. 
HB inference allows information to flow between the different data sets leading to more robust and accurate predictions for the model parameters.
These inferences are denoted $HB_{12, R}$ and $HB_{\pp, R}$. We use the experimentally measured RDFs from Ref.~\cite{Eisenstein:1942} as calibration data for these inferences. The RDFs are computed for 6 temperature/pressure pairs $(T, P)$. We denote the pairs as follows: $L_1 = (84.4, 0.8)$, $L_2 = (91.8, 1.8)$, $L_3 = (126.7, 18.3)$, $L_4 = (144.1, 37.7)$, $L_5 = (149.3, 46.8)$, $V = (149.3, 43.8)$, where $L$ stands for ``liquid'' and $V$ stands for ``vapor''. The corresponding datasets (RDFs) are denoted as $R_{Li}$ for liquid and $R_V$ for vapor. Finally, we perform the inference with LJ 6-$\pp$ using quantum dimer energy calculations from Ref.~\cite{Halpern:2010} as data and compare the obtained parameter distributions with those computed from the RDF data. The quantum dimer dataset is denoted as $Q$ and the corresponding inference is denoted as $B_{\pp, Q}$.

\subsection{Calibration of LJ 6-12}
\begin{figure}[b!]
    \centering
    \includegraphics[width=\textwidth]{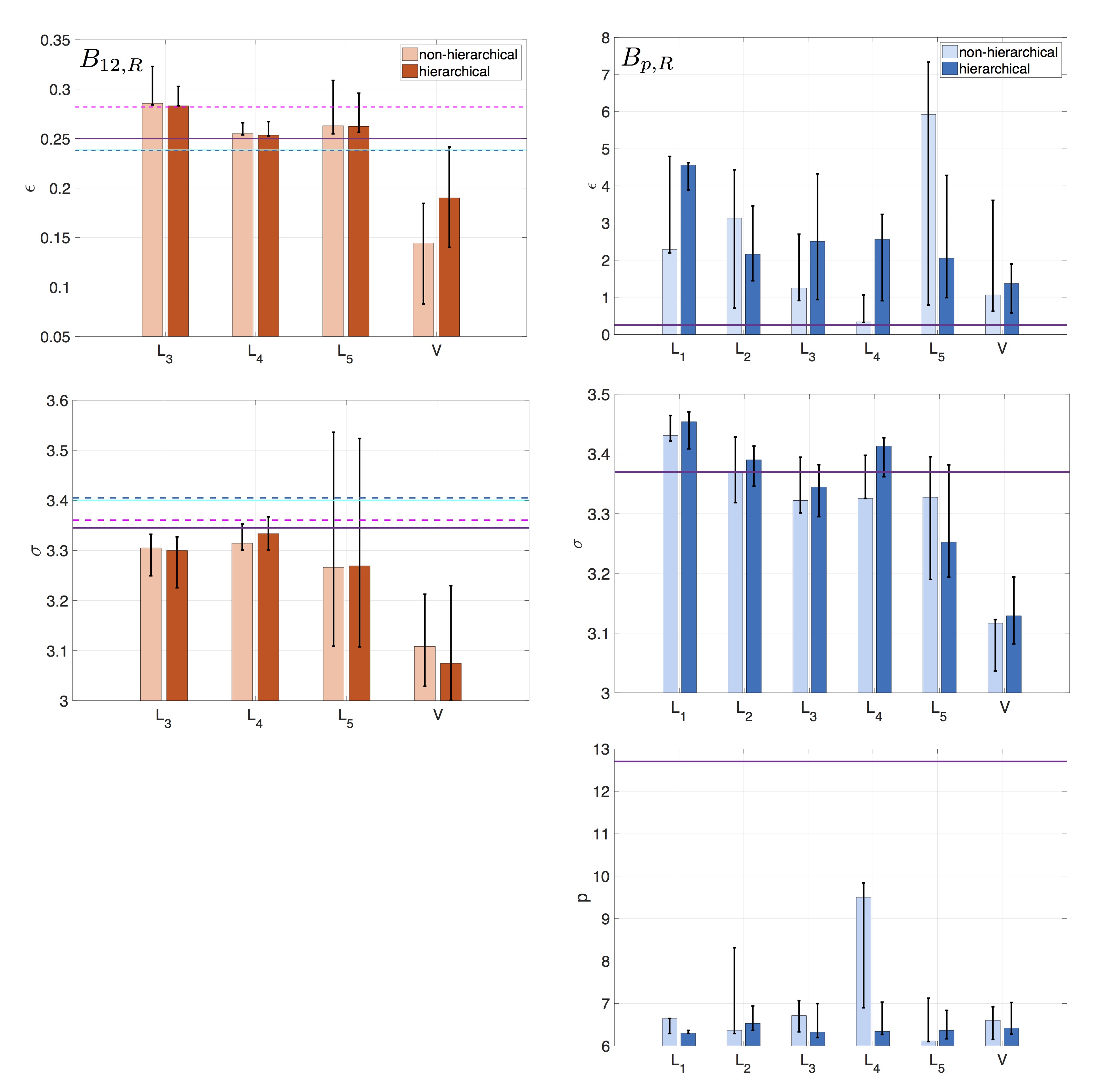}
    \caption{Posterior parameter values: MPVs along with 5\%-95\% quantiles obtained in $B_{12, R}$ (light red), $HB_{12, R}$ (dark red), $B_{p,R}$ (light blue), $HB_{p,R}$ (dark blue). Horizontal lines for LJ 6-12 indicate the reference values: Ref.~\cite{Barker:1971} (magenta dashed line), Ref.~\cite{White:1999} (purple solid line), Ref.~\cite{Rahman:1964} (cyan solid line), Ref.~\cite{Rowley:1975} (blue dashed line). Horizontal line for LJ 6-$p$ indicates the MPV for $B_{p,Q}$.}
    \label{fig:params}
\end{figure}

We present results of parameter calibration for $\pe, \ps, \pn$, while $\pp$ is fixed to 12. We use a wide enough uniform prior for each of the parameters and each of the datasets $R_{Li}$, $R_V$ ($\vp \in [0.05, 3] \times [3, 4] \times [10^{-6}, 1]$). We observe that the values which were obtained in the calibration process are close to those found in literature. In Fig.~\ref{fig:params} the MPVs of the parameters along with 5\%-95\% quantiles is presented (light red). Notice that results for the four out of six datasets are only presented since the LJ 6-12 potential failed to simulate the liquid argon for conditions $L_1$ and $L_2$.
A large difference in the values of $\pe$ for liquid and vapor is observed, which implies that one cannot perform the simulations using the same parameters for the two phases. We define the uncertainty in a parameter as the ratio of the 5\%-95\% quantile spread to the most probable value (MPV). The uncertainty in $\pe$  varies from 14\% to 20\% depending on the dataset, while the value of $\ps$ is identified more precisely with  uncertainty of 2\%-6\%. This difference can be attributed to the type of data used in the inference process: the location of the RDF peak,  which gives the most significant contribution to the sum of squared errors (SSE) inside the log-likelihood, is more sensitive to $\ps$. On the other hand, $\pe$ affects the height of the RDF peak which has a smaller effect on the log-likelihood.

Next, we infer the LJ parameters using the HB approach. We select the prior  $\pr{\vp_i}[\vhp]$ by using Bayesian model selection (see~\smref{sec:hb_info} for details). 
The values of the LJ parameters are presented in Fig.~\ref{fig:params} (dark red). The MPVs and the quantiles of the parameters are almost the same as in the $B_{12, R}$, which means that for each dataset $\vy_i \in \{R_{L1}, \ldots, R_{L5}, R_V\}$ no information about the parameters can be extracted from the other datasets.

\begin{figure*}[hbtp]
    \centering
    \includegraphics[width=\textwidth]{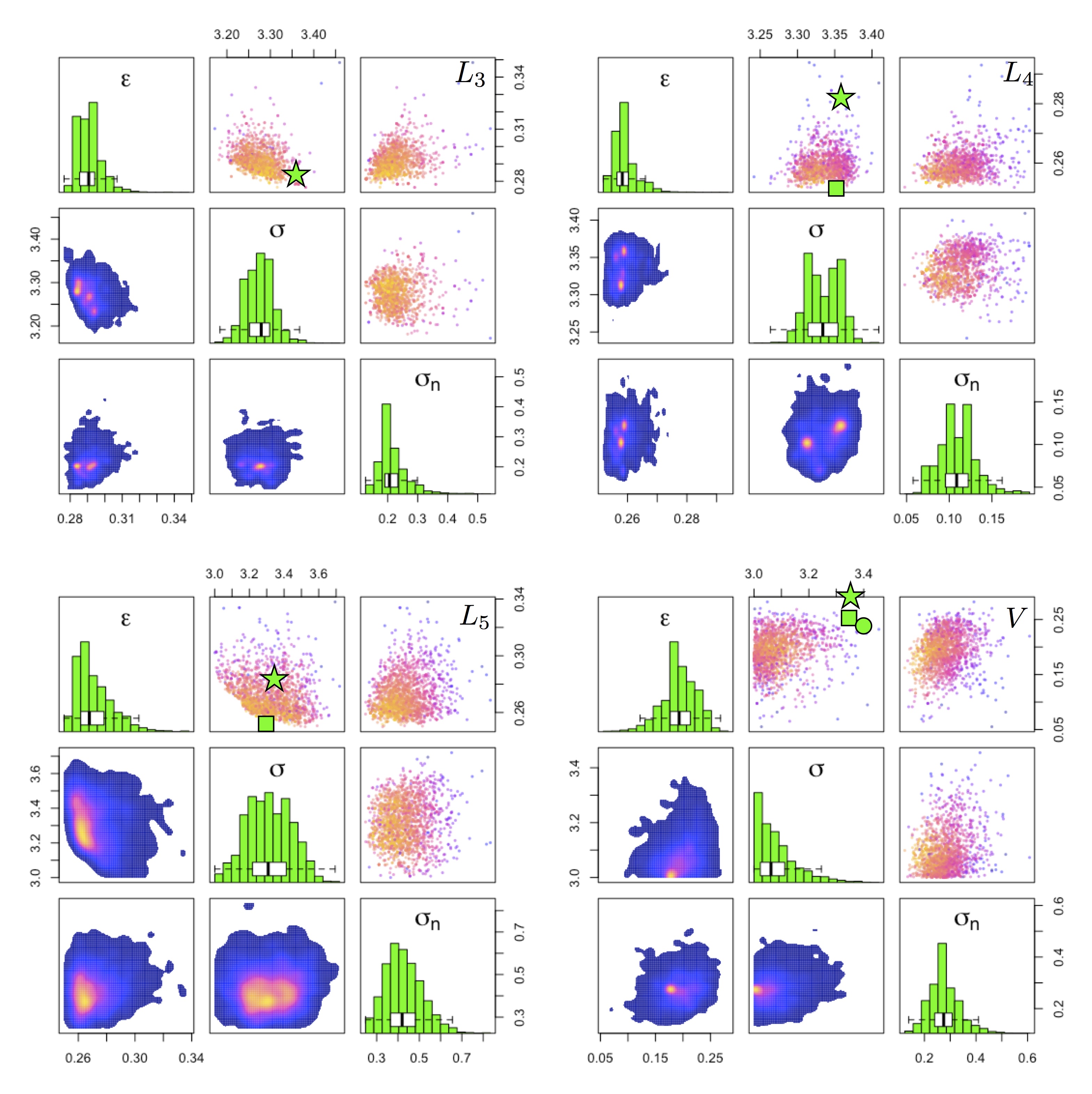}
    \caption{LJ parameters distributions obtained in $HB_{12, R}$ for $L_3$ (top left), $L_4$ (top right), $L_5$ (bottom left) and $V$ (bottom right).}
    \label{fig:pdf_12}
\end{figure*}

The full set of the MPVs and distribution quantiles for each dataset $R_{Li}$, $R_V$ is given in Table~\ref{tab:posterior_12}, while the full posterior distributions are shown in Fig.~\ref{fig:pdf_12}.

\subsection{Calibration of LJ 6-$\pp$}
\paragraph{Dataset $R$:} Here we include the LJ exponent $\pp$ into the parameter set $\vp$. As in the LJ 6-12 case, we choose a uniform prior with wide enough bounds ($[0.05, 10] \times [3, 4] \times [6.01, 15] \times [10^{-6}, 1]$). Note that with LJ 6-$\pp$ the sampling algorithm dictated much wider bounds for $\pe$ compared to the LJ 6-12 case. As will be seen later, this is due to a strong correlation between $\pe$ and $\pp$. We observe again the non-transferability of the LJ parameters from liquid to vapor simulations: the values of $\ps$ lie in disjoint domains for $L_i$ and $V$ (Fig.~\ref{fig:params}). Being a more flexible potential, LJ 6-$\pp$ can simulate a wider range of thermodynamic conditions, including $L_1$ and $L_2$, which result in the values of LJ parameters similar to those obtained for the other three liquid conditions. We observe that the 95\% quantile of $\pp$, as well as its MPV, is for four out of six RDF datasets below $7.5$ and for all the datasets below 10, which is much smaller than the conventional 12. This can be explained by the fact that the repulsion energies predicted by the standard 12-6 LJ potential are very high for the liquids. The configurations with such energies happen with probability close to zero, and the MD simulation is not able to sample them. As in the LJ 6-12 case, the parameter $\pe$ exhibits significant variation within each dataset $R_{Li}$, $R_V$ (uncertainty 110\%-216\%, computed the same way as for LJ 6-12), while $\pp$ and $\ps$ are well-defined with the uncertainty of 5\%-30\% and 1\%-6\%, respectively. In addition, $\pe$ differs substantially among the RDF datasets, but always in accordance with $\pp$: the higher the $\pp$, the lower the $\pe$ (see Fig.~\ref{fig:p_eps_all}).

\begin{figure*}[b!]
    \centering
    \includegraphics[width=\textwidth]{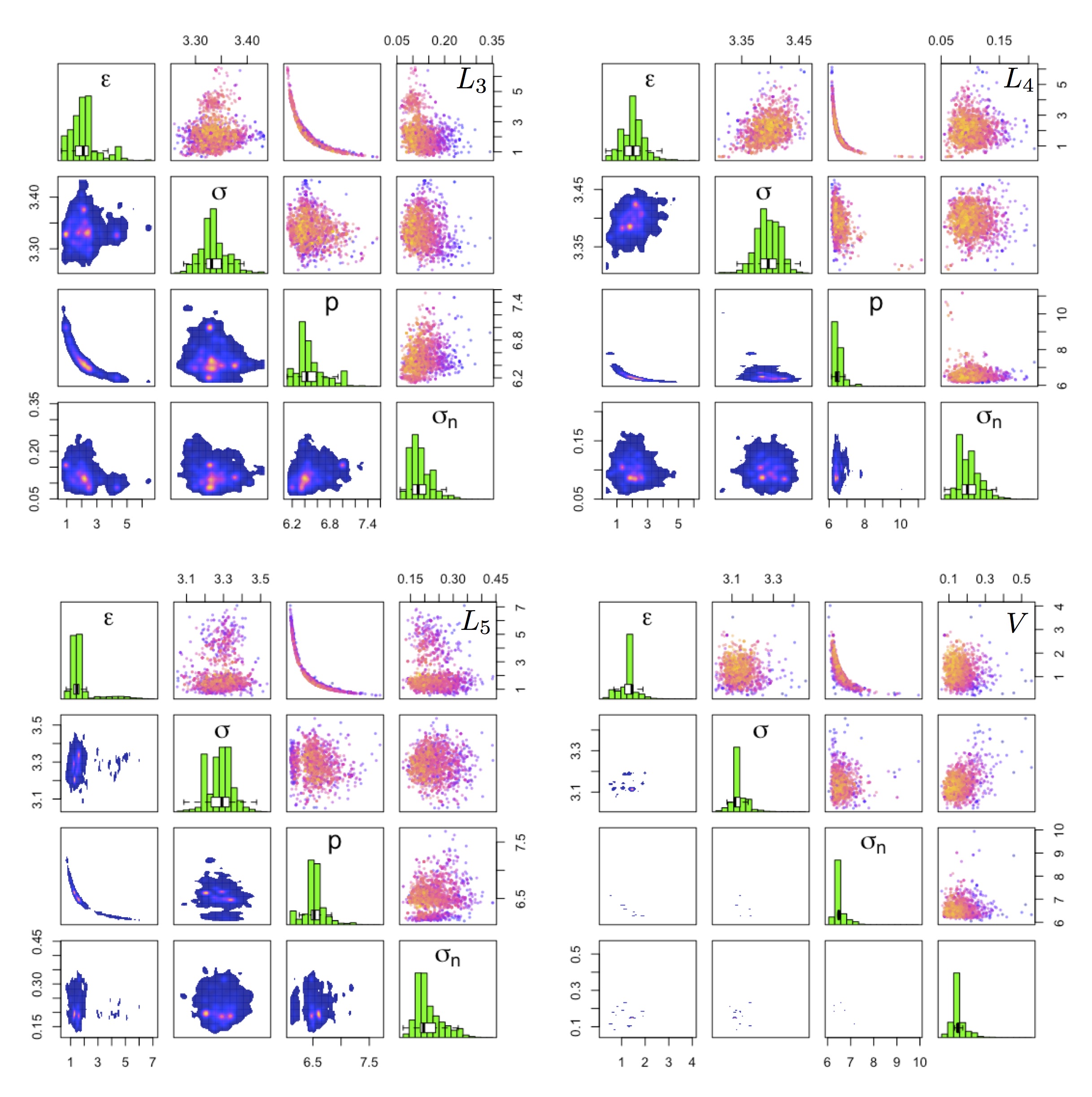}
    \caption{LJ parameters distributions obtained in $HB_{\pp, R}$ for $L_3$ (top left), $L_4$ (top right), $L_5$ (bottom left) and $V$ (bottom right).}
    \label{fig:pdf_free2}
\end{figure*}

\begin{figure*}[hbtp]
    \centering
    \includegraphics[width=\textwidth]{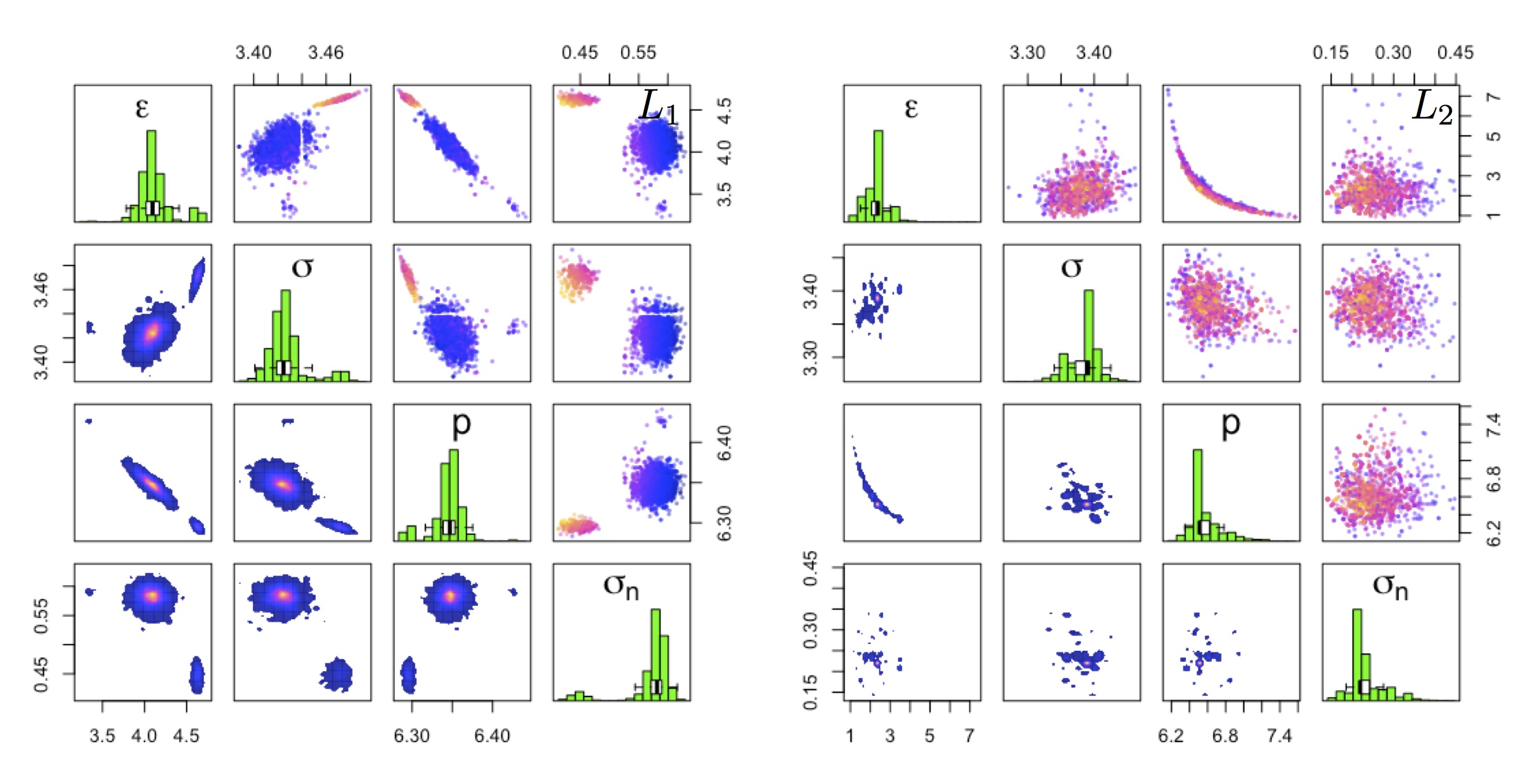}
    \caption{LJ parameters distributions obtained in $HB_{\pp, R}$ for $L_1$ (left) and $L_2$ (right).}
    \label{fig:pdf_free1}
\end{figure*}

Similarly with the inference of the LJ 6-12 parameters, we proceed by calibrating the parameters using the HB approach. Details for the section of the prior can be found in~\smref{sec:hb_info}.
The results of the inference are given in Fig.~\ref{fig:params}. {We observe that the uncertainty in $\pe$ gets significantly reduced for conditions $L_1$, $L_2$, $L_5$ and $V$ indicating that the inference benefited from the information contained in the two remaining datasets $R_{L3}$ and $R_{L4}$ with narrow posterior distributions of $\pe$ (Fig.~\ref{fig:pdf_free1}, \ref{fig:pdf_free2}). On the other hand, the uncertainty in $\pe$ for $L_3$ and $L_4$ increases adjusting to the wide ranges in the other four cases. A similar situation can be seen for $\pp$, where narrow distributions for $L_1$, $L_3$, $L_5$, $V$ shift the posterior values for $L_2$ and $L_4$. The RDF is, as noticed before, very sensitive to the changes in $\ps$, which controls the location of the LJ potential well, and therefore $\ps$ is well determined for each of the datasets $R_{L_i}$, $R_V$ and extracts almost no information from the other ones.}

\paragraph{Dataset $Q$:} To investigate whether the repulsion exponent 12 may be a good model for some cases, we perform a calibration using the calculated quantum dimer scans of argon as data. These data describe the behavior of the gaseous argon. We infer the LJ 6-$\pp$ parameters by fitting the LJ potential to the binding energy of the quantum dimer (Fig.~\ref{fig:params}). The resulting value of $\pp$ is much closer to the conventional 12 (Table~\ref{tab:posterior_free}), suggesting that for the gaseous argon, unlike for the liquid one, LJ 6-12 is a reasonable choice.

The full set of MPVs and distribution quantiles of the LJ parameters for $R_{Li}$, $R_V$, $Q$ is given in Table~\ref{tab:posterior_free} and the full posterior distributions are plotted in Fig.~\ref{fig:pdf_free1}, \ref{fig:pdf_free2}, \ref{fig:pdf_quantum}. In Section~\ref{sec:comparison}

\subsection{Experimental Data vs Quantum Mechanics Simulations: Model Comparison} \label{sec:comparison}

\paragraph{Model selection} We select between LJ 6-12 and LJ 6-$\pp$ potentials by applying the Bayes selection criterion. We observe that LJ 6-$\pp$ is significantly better than the LJ 6-12 for $L_3$ and $L_5$ (Table~\ref{tab:model_selection}). Recalling that LJ 6-12 is not able to produce a liquid for $L_1$ and $L_2$, we conclude that LJ 6-$\pp$ is preferred for four RDF datasets out of six. In the case of $L_4$ the potentials show indistinguishable by the Bayesian model selection results. The only dataset on which the LJ 6-12 potential produces better results (3 times more probable than LJ 6-$\pp$) is $V$, the vapor case. That brings us to the conclusion that LJ 6-$\pp$ is either much better or not worse than LJ 6-12 for all the liquid cases considered. For the vapor case, the LJ 6-$\pp$ is over parametrized, as compared to LJ 6-12.

\begin{figure}[bt]
    \centering
    \includegraphics[width=0.45\textwidth]{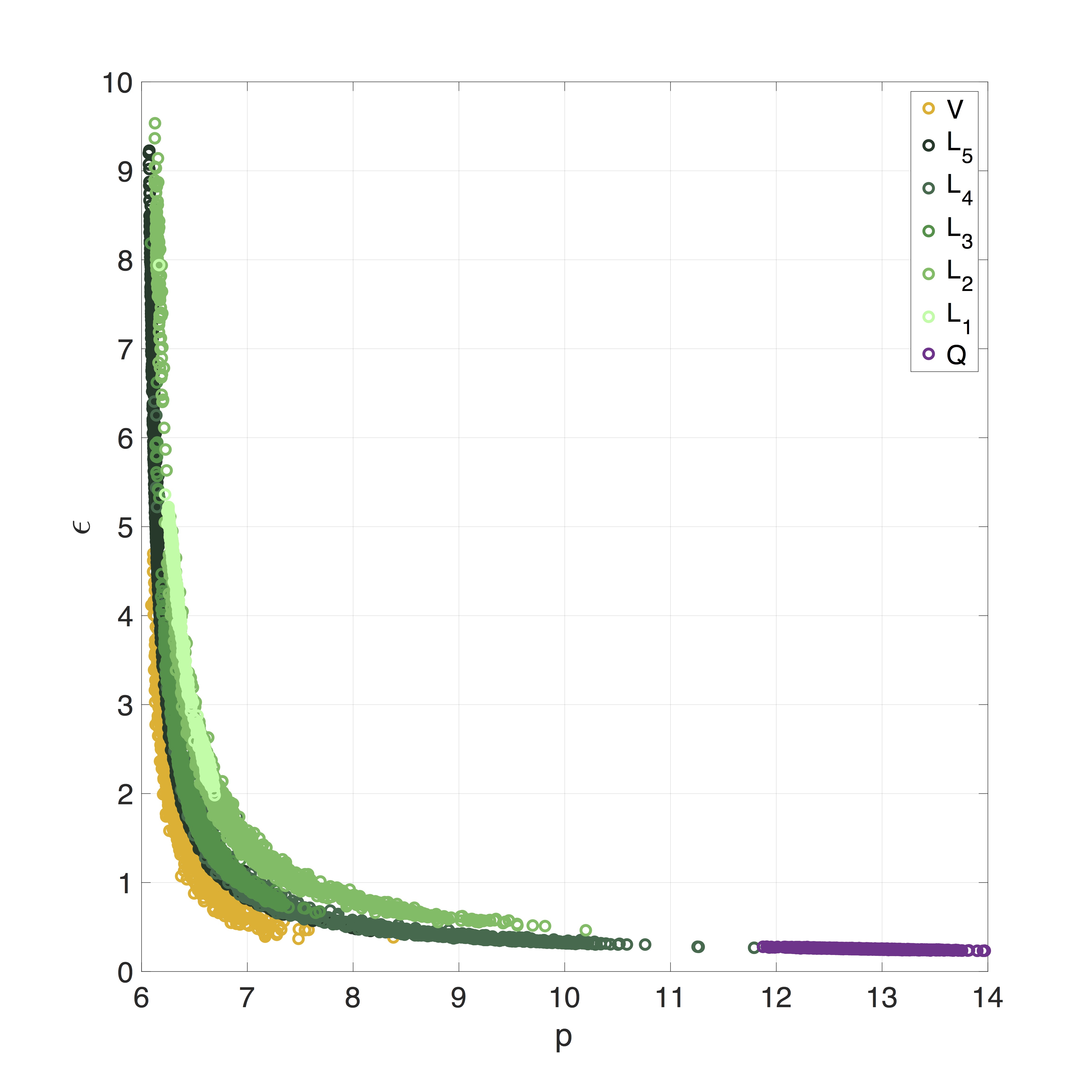}
    \caption{Posterior samples of $B_{\pp, R}$ projected onto $(\pp, \pe)$ subspace: yellow circles correspond to $V$, green circles correspond to $L_i$ in the temperature increasing order from the lightest to the darkest color, purple circles correspond to $Q$.}
    \label{fig:p_eps_all}
\end{figure}

\begin{figure*}[bt]
    \includegraphics[width=\textwidth]{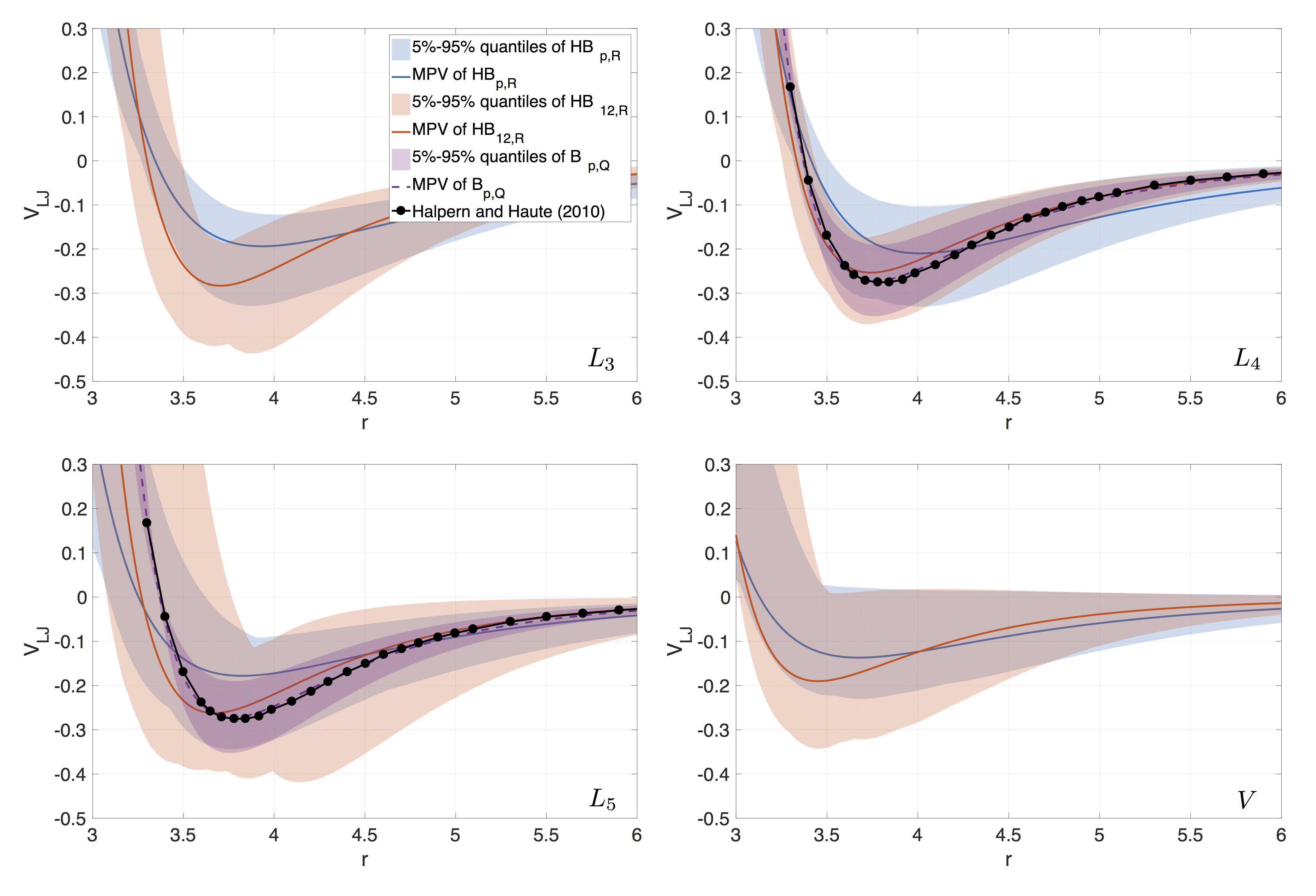}
    \caption{Posterior LJ potentials comparison: MPVs along with 5\%-95\% quantiles obtained in $HB_{\pp, R}$ (blue), $HB_{12, R}$ (red), $B_{\pp, Q}$ (purple). Black line with dots: quantum dimer calculations.}
    \label{fig:LJ_12_free}
\end{figure*}

\begin{figure}[hbtp]
	\begin{minipage}{0.5\textwidth}
		\centering
		\includegraphics[width=\textwidth]{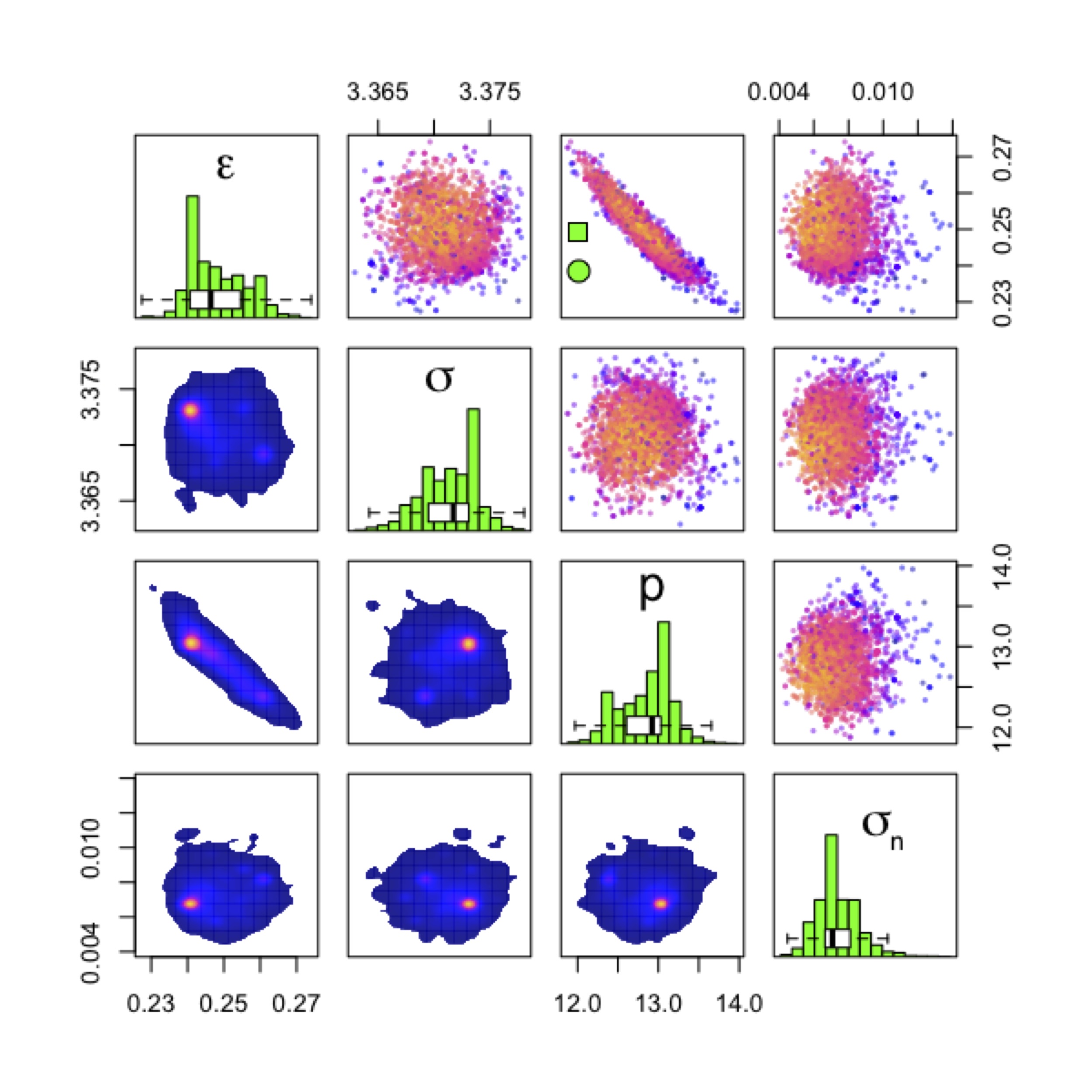}
		\vspace{-0.8cm}
		\caption{LJ parameters distributions obtained in $B_{\pp, Q}$.}
		\label{fig:pdf_quantum}
	\end{minipage}
	\begin{minipage}{0.5\textwidth}
		\centering
		\includegraphics[width=\textwidth]{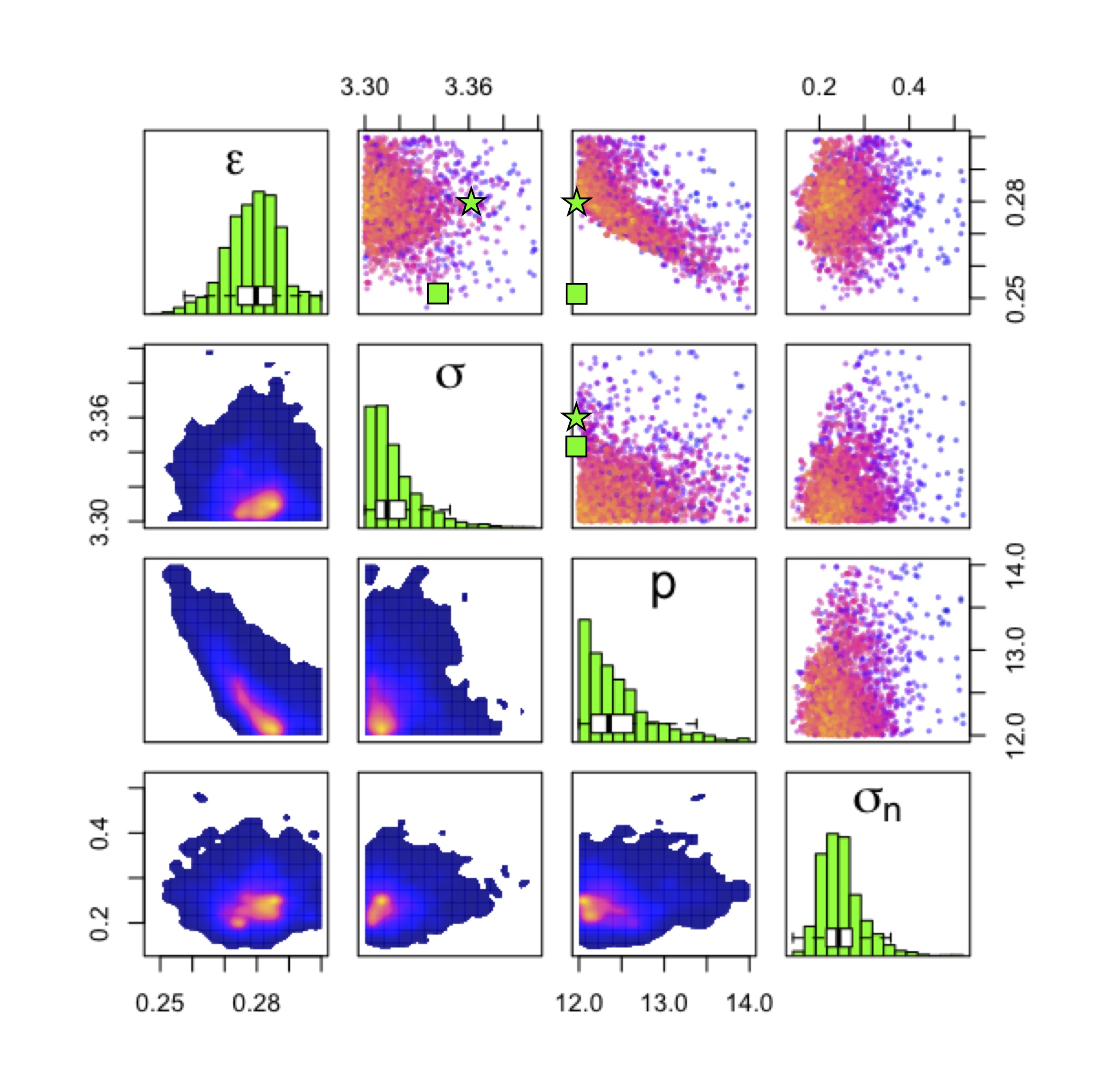}
		\vspace{-0.8cm}
		\caption{LJ parameters distributions obtained in $B_{\pp, R}$ for $L_3$ with the prior for $\pp$ restricted to $[12, 14]$.}
		\label{fig:pdf_126_narrow}
	\end{minipage}
\end{figure}

\paragraph{LJ potentials} Studying the reasons for LJ 6-$\pp$ being more plausible than LJ 6-12, we take a closer look at the inferred shapes of the potentials. We observe a very stable correlation in the $(\pp, \pe)$ subspace (Fig.~\ref{fig:p_eps_all}) for all the datasets used.
This result is expected as $\pp$ regulates the strength of the repulsion and $\pe$ alters the strength of both repulsion and attraction simultaneously. The difference between the $R_i$ and $Q$ datasets shows up in the region of the subspace which gets populated. The quantum dimer-based calibration prefers high values of $\pp$, which correspond to the tails of the distributions inferred using the RDF data. We performed a calibration with $L_3$ and narrow prior bounds ($\pp \in [12, 14]$) to see whether this is indeed a tail of the full posterior distribution (Fig.~\ref{fig:pdf_126_narrow}). The narrow posterior values are below $3.95$, while the values of the full posterior start from $4.18$, which explains why the tails of the full distributions for $L_i$, $V$ have a negligible number of samples in the region $\pp \in [12, 14]$ preferred by the $Q$-based inference. As the parameters $\pe$ and $\pp$ are highly correlated, one could expect that the inference will be able to recover values of $\pe$ for LJ 6-12 such that the resulting potential is close to the inferred LJ 6-$\pp$. However, the effect that $\pp$ and $\pe$ have on the LJ potential is not entirely the same. As $\pe$ acts as a scaling factor for the whole potential, it is not able to make the potential less deep and at the same time flat enough to avoid switching to the gas phase (compare simulations with MPVs for $L_5$, $V$ in Fig.~\ref{fig:LJ_12_free}). The same reasoning can be applied to explain the inability of LJ 6-12 to drive $L_1$ and $L_2$ to the liquid phase: the potential is too repulsive, frustrating the liquid packing, and the system behaves either like a gas or like a solid (note that $L_1$ is close to the argon triple point).

The full set of the inferred LJ 6-12 and LJ 6-$\pp$ potentials is given in Fig.~\ref{fig:LJ_12_free}.

\paragraph{Robust posterior prediction} The quality of the predictions made for a QoI different than the one used for the inference quantifies the predictive power of the model (see~\smref{sec:prediction}). Making predictions for new QoIs is a challenging problem in MD as each quantity depends on LJ parameters in a specific non-linear fashion. We obtain robust predictions of the RDF, density $\rho$ and diffusion coefficient $D$ of argon by propagating the posterior LJ parameters uncertainty into these quantities.

We measure the error $\Delta q$ of the prediction of the scalar quantity $q$ as
\begin{equation}
    \Delta q = \frac{1}{N} \sum_{k=1}^N \lb \frac{q_k-r_k}{r_k} \rb^2 \cnt
\end{equation}
where $N \leq 6$ is the number of thermodynamic conditions for which the prediction can be made, $q_k$ is the prediction made using the MPV and $r_k$ is the reference value. The reference values for $\rho$ are experimental measurements taken from Ref.~\cite{Eisenstein:1942}. The reference values for $D$ are computed analytically using the equations from Ref.~\cite{Kestin:1984}. The accuracy of the fit for these computations is 0.7\%. The error of the RDF is computed as an average over all the thermodynamic conditions mean squared error of the computed RDF vs the experimental RDF. The predictions are compared on three different sets of conditions: 1) the conditions which can be simulated using MPVs obtained in all the three inferences $HB_{12, R}$, $HB_{\pp, R}$, and $B_{\pp,Q}$ ($L_4$, $L_5$), 2) the conditions which can be simulated using MPVs obtained in the inferences $HB_{12, R}$ and $HB_{\pp, R}$ ($L_3-L_5$, $V$), 3) the conditions which can be simulated using MPVs obtained in the inference $HB_{\pp, R}$ ($L_1-L_5$, $V$).

The predictions made using the results of $HB_{\pp, R}$ are the most accurate for all the QoIs considered and all the sets of conditions, except for one case where $HB_{12,R}$ gives a better result (see Table~\ref{tab:fp}). On the other hand, the predictions made using the results of $B_{\pp,Q}$ are the least accurate for all the QoIs. Additionally, the inferences $HB_{12,R}$ and $B_{\pp,Q}$ result in LJ potentials which cannot be used to simulate all the thermodynamic conditions. This brings us to the conclusions that 1) $HB_{\pp,R}$ produces a better LJ model than $HB_{12,R}$, 2) $B_{\pp,Q}$  does not result in a good model for liquid argon or saturated argon vapor.

We note that the values of $D$ differ by an order of magnitude for liquid and vapor which explains the huge deterioration of the predictions on the sets of conditions that include $V$.

\begin{figure*}[tp]
    \centering
    \includegraphics[width=\textwidth]{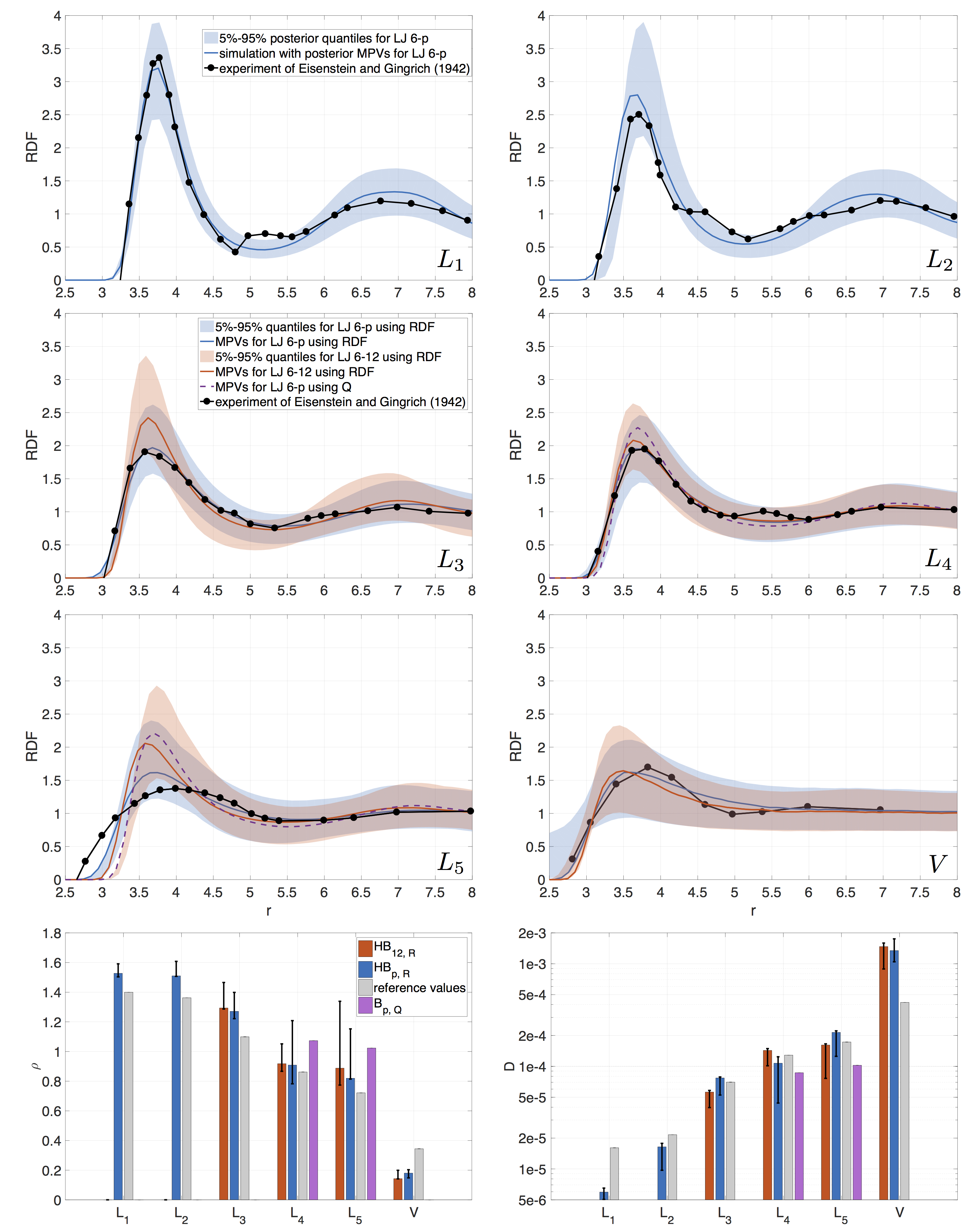}
    \caption{Robust posterior predictions: MPVs along with 5\%-95\% quantiles obtained in $HB_{\pp, R}$ (blue), $HB_{12, R}$ (red), MPV obtained in $B_{\pp, Q}$ (purple). Black line with dots: experimental data for RDF. Grey bars: experimental data for $\rho$, analytically computed values for $D$.}
    \label{fig:fp}
\end{figure*}

The MPVs of $D$ and $\rho$ along with the corresponding quantiles are presented in Fig.~\ref{fig:fp}. The same values for RDF are given in Fig.~\ref{fig:fp}.
%
%%%%%%%%%%%%%%%%%%%%%%%%%%%%%% END RESULTS %%%%%%%%%%%%%%%%%%%%%%%%%%%%%%%%%%%%%

\section{Discussion} \label{sec:disc}
%%%%%%%%%%%%%%%%%%%%%%%%%%%%%% DISCUSSION %%%%%%%%%%%%%%%%%%%%%%%%%%%%%%%%%%%%%%
We examine the classical  6-12 Lennard Jones potential using Hierarchical Bayesian inference with data form experiments and quantum mechanics simulations. 
Our results show that the value ($p=12$) of the repulsive exponent needs to be revised and in the case of argon be replaced by a smaller value ($p=6.5$).
Notably we find that calibration for the repulsive exponent is more accurate and robust when using  experimental data rather than  data from quantum mechanics simulations  The results  indicate that  parameters inferred from the quantum dimer calculations are not predictive for the liquid and saturated vapor conditions and that  smaller values of the exponent $\pp$ ($\pp \in (6,9)$) in the Lennard-Jones potential provide better predictions for RDF (Fig.~\ref{fig:fp}), density and diffusion data (Fig.~\ref{fig:fp}) than the conventional $\pp=12$ or $\pp=12.7$ inferred from $Q$. These new  LJ exponents allow to simulate a larger variety of thermodynamic conditions but  cannot be transferred from liquid to gas using this simplified model (Fig.~\ref{fig:params}). We have also  examined whether the smaller exponent allows for bigger time steps in MD simulations. However it appears that the exponent is not a critical factor for the stability of the system. We observed similar execution times for the simulations with MPVs of LJ 6-12 and LJ 6-$\pp$. At the same time usage of the surrogates resulted in a speed-up of 28\% for the LJ 6-12 case. For the LJ 6-$\pp$ case the unidentifiable manifold in the parameter space $(\pp, \pe)$ did not allow for an efficient kriging approximation. Our  results contradict the conclusion of Ref.~\cite{Galliero:2006}, where LJ potentials with $\pp=10, 12, 14, 16, 18, 20$ were fit to viscosity and pressure data, and the potential with $\pp=12$ showed better agreement for different thermodynamic conditions. This mismatch can be explained by the fact that different data was used and also that the exponents below 10, which appear to be the best according to the results of the current study, were not tested in Ref.~\cite{Galliero:2006}. 
The present results suggest that experimental data are more suitable for robust predictions in calibrated MD potentials and suggest that similar studies are necessary across all fields that employ MD simulations.
%%%%%%%%%%%%%%%%%%%%%%%%%%%%%% END DISCUSSION %%%%%%%%%%%%%%%%%%%%%%%%%%%%%%%%%%

\section{Methods} \label{sec:methods}
%%%%%%%%%%%%%%%%%%%%%%%%%%%%%% METHODS %%%%%%%%%%%%%%%%%%%%%%%%%%%%%%%%%%%%%%%%%
\subsection{Molecular Dynamics} \label{sec:sim}
We perform MD simulations of argon using LAMMPS package~\cite{lammps}. The argon atoms are modeled as spheres which interact with LJ 6-$\pp$ potential:
\begin{equation}
	V_{LJ}(r; \pe, \ps, \pp) = 4\pe \lb \lb \frac{\ps}{r} \rb^{\pp} - \lb \frac{\ps}{r} \rb^6 \rb \cnt
\end{equation}
where $r$ is the distance between the interacting atoms and $\pp$ is the repulsion exponent usually taken to be 12. The parameters $\pe$, $\ps$ and $\pp$ are to be chosen according to the available measurements. As the Lennard-Jones interactions quickly decay with the distance, an additional computational parameter $r_c$ is usually introduced. This parameter defines a cut-off distance at which the potential is set to zero. Here, we set $r_c = 2.5\ps$. The thermodynamic state of the system is defined by the temperature and the pressure of the argon atoms. We ensure that argon is in the liquid/vapor state by checking the self-diffusion coefficient and the density. The simulation starts with energy minimization followed by $5 \times 10^{6}$ steps, of 2 fs, in an NPT ensemble. Then the RDF is computed in the production run consisting of $10^{5}$ NVE integration steps of 2 fs each. The boundary conditions are periodic in each direction, the domain contains 666 argon atoms. The self-diffusion coefficient is calculated via the mean-squared displacement of the atoms, the RDF is discredited using 100 bins. The units used in the current work are given in Table~\ref{tab:units}.

\subsection{Bayesian Uncertainty Quantification} \label{sec:uq_th}
This section presents a brief description of the Bayesian inference theory. The details are given in~\smref{sec:uq_th_si}. Here and further in the text small bold letters represent vectors while big bold letters represent matrices. Each random variable $\vec{\xi}$ is assumed to be continuous with a probability density function (PDF) denoted as $\pr{\vec{\xi}}$. 

Let $f(\vx; \vp) \in \R^{M}$ denote the output, or a quantity of interest (QoI), of a computational model with input $\vx \in \R^{N_x}$ and parameters $\vp = (\p_1, \ldots, \p_{N_{\p}}) \in \R^{N_{\p}}$. Let also $\vy \in \R^{N_{\y}}$ be a vector of experimental data corresponding to the QoI $f$ and input parameters $\vx$. The experimental data are linked with the computational model through the likelihood function, $\pr{\vy}[\vp, \vx]$. A usual model assumption for the likelihood function involves a Gaussian,
\begin{equation} \label{eq:likelihood}
    \pr{\vy}[\vp, \vx] = \N(\vy \,|\, f(\vx; \vp), \C) \cnt
\end{equation}
where $\C$ is a covariance matrix that may be a function of $\vp$. To simplify the notations, the conditioning on $\vx$ is omitted below. Prior information on the parameters $\vp$ is encoded into the probability distribution with PDF $\pr{\vp}[\M]$. We assume $\C = \pn^2 \vec{I}$, where $\vec{I}$ is the identity matrix in $\R^{N_{\p} \times N_{\p}}$ and $\pn \in \R$ is \textit{a priori} unknown. In this work, we infer the parameters of the LJ potential together with the parameter of the covariance matrix: $\vp = (\pe, \ps, \pn)$ or $\vp = (\pe, \ps, \pp, \pn)$ depending on whether that exponent $\pp$ is being inferred or not.

Bayes' theorem provides a tool for the inference of the parameters $\vp$ conditioned on the observations $\vy$,
\begin{equation}
    \pr{\vp}[\vy, \M] = \frac{ \pr{\vy}[\vp, \M] \, \pr{\vp}[\M] }{ \pr{\vy}[\M] } \cnt
\end{equation}
where $\pr{\vy}[\M] = \int \pr{\vy}[\vp, \M] \, \pr{\vp}[\M] d\vp$ is a normalization constant and $\M$ stands for ``model'', which is a set of the assumptions regarding the likelihood and the prior. We remark that the denominator $\pr{\vy}[\M]$, called model evidence, is used for model selection (see~\smref{sec:model:selection}).

In certain cases the data may correspond to different input variables $\vx$ of the model, one of the examples is pressure and temperature used in this work. Let $\vvy = \{\vy_1, \ldots, \vy_N\}$ be the set of all provided data with $\vy_i \in \R^{N_{\p_i}}$, where each $\vy_i$ corresponds to different input $\vx_i$. In this case one wishes to infer different parameters, $\vp_i \in \R^{N_\p}$, for each dataset $\vy_i$. Here, we assume that the parameters $\vp_i$ depend on hyper-parameters $\vhp \in \R^{N_\hp}$, which encode the variability of $\vp_i$ between the datasets and should also be inferred.

For the sampling of the distributions we use the Transitional Markov Chain Monte Carlo (TMCMC) algorithm~\cite{Ching:2007} (see~\smref{sec:sampling:posterior}).
We perform all the inferences using the open-source library $\Pi$4U~\cite{Hadjidoukas:2015a} on Brutus cluster of the ETH Zurich and Piz Daint cluster of the Swiss National Supercomputing Center (CSCS). We use 2000 samples per TMCMC stage for LJ 6-12 and 4000 samples per stage for LJ 6-$\pp$. The parallelisation is made with MPI and internal worker threads of the $\Pi$4U library. The task-based parallelism and the load balancing mechanisms of $\Pi$4U provide the necessary flexibility for running MD simulations with very different execution time within TMCMC.

In order to reduce the computational cost of the simulations, we apply kriging surrogates following the methodology proposed in Ref.~\cite{Angelikopoulos:2015}. Namely, for each Markov chain leader we build a kriging interpolating surface using the samples from the leader's bounding box. We select the size of the box to be equal to a quarter of the current domain. The surrogate value is rejected if the kriging error is greater than 5\% of the predicted value. In addition, we do not allow the kriging predictions which are outside the 5\%-95\% quantile range of all the values obtained from MD simulations.
%%%%%%%%%%%%%%%%%%%%%%%%%%%%%% END METHODS %%%%%%%%%%%%%%%%%%%%%%%%%%%%%%%%%%%%%

%%%%%%%%%%%%%%%%%%%%%%%%%%%%%% TABLES %%%%%%%%%%%%%%%%%%%%%%%%%%%%%%%%%%%%%%%%%%
%
\begin{table*}[hbtp]
	\caption{Units used in this work.}
	\label{tab:units}
	\centering
	\begin{tabular}{l l l}
		\hline \hline
		name & notation & \textit{real} \\
		\hline
		Temperature                & $T$    & K        \\
		Pressure                   & $P$    & atm      \\
		Distance                   & $r$    & \AA      \\
		LJ well depth              & $\pe$  & kcal/mol \\
		LJ well location           & $\ps$  & \AA      \\
		LJ repulsion exponent      & $\pp$  & --       \\
		RDF model error            & $\pn$  & --       \\
		Density                    & $\rho$ & g/cm$^3$ \\
		Diffusion coefficient      & $D$    & cm$^2$/s \\
		\hline \hline
	\end{tabular}
\end{table*}
\begin{table*}[hbtp]
	\caption{LJ parameters for argon used in literature. The last row shows the data used for fitting. Notations: $T$ (temperature), $P$ (pressure), $\rho$ (density), $B$ (second virial coefficient), $E$ (energy), $TP$ (gas-liquid transition pressure), $L$ (latent heat of evaporation).}
	\label{tab:LJ}
	\centering
	\begin{tabular}{l l l l l}
		\hline \hline
		& Ref.~\cite{Rahman:1964} & Ref.~\cite{Barker:1971} & Ref.~\cite{Rowley:1975} & Ref.~\cite{White:1999}   \\
		\hline
		$\pe$  & 0.2385 & 0.2824             & 0.2381            & 0.2498        \\
		$\ps$  & 3.4000 & 3.3605             & 3.4050            & 3.3450        \\
		$T$    & 94.4   & 86.64 - 168.86     & 137.77            & 88 - 127      \\
		$\rho$ & 1.374  & 0.435 - 1.479      & 0.156, 0.972      & 0.283 - 3.897 \\
		Phase  & liquid & gas, liquid, solid & gas + liquid      & gas + liquid  \\
		Data   & RDF    & $P$, $E$           & $TP$, $\rho$, $L$ & $P$, $B$      \\
		\hline \hline
	\end{tabular}
\end{table*}
\begin{table*}[hbtp]
	\caption{Posterior values of each parameter $\p \in \{\pe, \ps, \pn\}$ of LJ 6-12: MPV $b(\p)$ and 5\%-95\% quantiles $q(\p)$.}
	\label{tab:posterior_12}
	\centering
	\begin{tabular}{l l l l l l l l}
		\hline \hline
		& & $b(\pe)$ & $q(\pe)$ & $b(\ps)$ & $q(\ps)$ & $b(\pn)$ & $q(\pn)$ \\
		\hline
		$B_{12, R}$  & $L_3$ & 0.286 & [0.284, 0.323] & 3.305 & [3.250, 3.332] & 0.168 & [0.158, 0.314] \\
					 & $L_4$ & 0.255 & [0.254, 0.266] & 3.314 & [3.301, 3.353] & 0.089 & [0.080, 0.140] \\
					 & $L_5$ & 0.263 & [0.255, 0.309] & 3.266 & [3.110, 3.536] & 0.317 & [0.292, 0.586] \\
					 & $V$   & 0.144 & [0.083, 0.184] & 3.109 & [3.029, 3.213] & 0.147 & [0.119, 0.304] \\
		\hline
		$HB_{12, R}$ & $L_3$ & 0.283 & [0.283, 0.303] & 3.300 & [3.226, 3.327] & 0.177 & [0.156, 0.310] \\
					 & $L_4$ & 0.253 & [0.254, 0.267] & 3.333 & [3.301, 3.367] & 0.073 & [0.073, 0.147] \\ 
					 & $L_5$ & 0.262 & [0.256, 0.296] & 3.269 & [3.108, 3.523] & 0.337 & [0.301, 0.579] \\
					 & $V$   & 0.190 & [0.140, 0.242] & 3.075 & [3.001, 3.229] & 0.180 & [0.192, 0.385] \\
		\hline \hline
	\end{tabular}
\end{table*}
\begin{table*}[hbtp]
    \caption{Posterior values of each parameter $\p \in \{\pe, \ps, \pp, \pn\}$ of LJ 6-$\pp$: MPV $b(\p)$ and 5\%-95\% quantiles $q(\p)$.}
    \label{tab:posterior_free}
    \centering
    \begin{tabular}{l l l l l l l l l l}
        \hline \hline
        & & $b(\pe)$ & $q(\pe)$ & $b(\ps)$ & $q(\ps)$ & $b(\pp)$ & $q(\pp)$ & $b(\pn)$ & $q(\pn)$ \\
        \hline
        $B_{\pp, R}$  & $L_1$ & 2.286 & [2.193, 4.794] & 3.431 & [3.422, 3.464] & 6.644  & [6.292,  6.647]  & 0.157 & [0.185, 0.302] \\
                      & $L_2$ & 3.134 & [0.712, 4.416] & 3.369 & [3.319, 3.428] & 6.370  & [6.293,  8.313]  & 0.222 & [0.215, 0.518] \\
                      & $L_3$ & 1.250 & [0.914, 2.700] & 3.322 & [3.301, 3.395] & 6.715  & [6.332,  7.068]  & 0.098 & [0.080, 0.159] \\
                      & $L_4$ & 0.337 & [0.322, 1.060] & 3.325 & [3.325, 3.398] & 9.501  & [6.900,  9.842]  & 0.057 & [0.066, 0.153] \\
                      & $L_5$ & 5.928 & [0.794, 7.318] & 3.328 & [3.190, 3.395] & 6.116  & [6.102,  7.126]  & 0.164 & [0.168, 0.326] \\
                      & $V$   & 1.065 & [0.625, 3.611] & 3.117 & [3.037, 3.122] & 6.604  & [6.151,  6.923]  & 0.100 & [0.099, 0.190] \\
        \hline
        $HB_{\pp, R}$ & $L_1$ & 4.561 & [3.890, 4.626] & 3.454 & [3.408, 3.471] & 6.302  & [6.296,  6.366]  & 0.422 & [0.450, 0.602] \\
                      & $L_2$ & 2.081 & [1.333, 3.515] & 3.387 & [3.335, 3.424] & 6.565  & [6.340,  7.051]  & 0.211 & [0.178, 0.351] \\
                      & $L_3$ & 2.506 & [0.941, 4.325] & 3.345 & [3.295, 3.382] & 6.324  & [6.198,  6.996]  & 0.093 & [0.081, 0.186] \\
                      & $L_4$ & 2.588 & [0.892, 3.676] & 3.403 & [3.358, 3.432] & 6.339  & [6.226,  7.063]  & 0.082 & [0.075, 0.146] \\
                      & $L_5$ & 2.055 & [0.992, 4.281] & 3.252 & [3.194, 3.382] & 6.364  & [6.169,  6.837]  & 0.183 & [0.159, 0.316] \\
                      & $V$   & 1.371 & [0.582, 1.896] & 3.129 & [3.082, 3.194] & 6.422  & [6.277,  7.025]  & 0.111 & [0.103, 0.233] \\
        \hline
        $B_{\pp, Q}$  &       & 0.252 & [0.239, 0.261] & 3.370 & [3.367, 3.375] & 12.703 & [12.333, 13.309] & 0.006 & [0.006, 0.010] \\ 
        \hline \hline
    \end{tabular}
\end{table*}
\begin{table*}[hbtp]
	\caption{Log-evidences $E_{12,R}$ ($E_{p,R}$) for $B_{12,R}$ ($B_{p,R}$).}
	\label{tab:model_selection}
	\centering
	\begin{tabular}{l r r l}
		\hline \hline
		& $E_{12,R}$ & $E_{p,R}$ & $e^{E_{p,R}-E_{12,R}}$ \\
		\hline
		$L_1$ &  --    &  -7.05 & -- \\
		$L_2$ &  --    &  -14.8 & -- \\
		$L_3$ &  -9.72 &   2.81 & 2.74$\times$10$^5$ \\
		$L_4$ &   5.10 &   5.18 & 1.09 \\
		$L_5$ &  -15.8 &  -8.76 & 1.18$\times$10$^3$ \\
		$V$   &  -3.83 &  -4.94 & 3.31$\times$10$^{-1}$ \\
		\hline \hline
	\end{tabular}
\end{table*}
\begin{table*}[hbtp]
	\caption{Errors of robust posterior predictions of RDF, density and diffusion coefficient using LJ 6-12 and LJ 6-$\pp$. We denote $S_1 = \{L_4, L_5\}$ (all inferences produce the correct argon phase), $S_2 = \{L_3 - L_5, V\}$ ($B_{\pp,Q}$ produces wrong phase), $S_3 = \{L_1 - L_5, V\}$ ($B_{\pp,Q}$ and $HB_{12,R}$ produce wrong phase).}
	\label{tab:fp}
	\centering
	\begin{tabular}{l c c c c c c c c c}
		\hline \hline
		\multicolumn{1}{c|}{} & \multicolumn{3}{c|}{$\Delta$ RDF} & \multicolumn{3}{c|}{$\Delta \rho$} & \multicolumn{3}{c}{$\Delta D$} \\
		\hline
		\multicolumn{1}{c|}{} & $S_1$ & $S_2$ & \multicolumn{1}{c|}{$S_3$} & $S_1$ & $S_2$ & \multicolumn{1}{c|}{$S_3$} & $S_1$ & $S_2$ & $S_3$ \\
		\hline
		$B_{\pp, Q}$  & 0.087 & --    & --    & 0.118 & --    & --     & 0.136 & --    & --    \\
		$HB_{12, R}$  & 0.071 & 0.050 & --    & 0.029 & 0.108 & --     & 0.009 & 1.573 & --    \\
		$HB_{\pp, R}$ & 0.016 & 0.024 & 0.027 & 0.011 & 0.068 & 0.049  & 0.043 & 1.230 & 0.898 \\
		\hline \hline
	\end{tabular}
\end{table*}
%
%%%%%%%%%%%%%%%%%%%%%%%%%% END TABLES %%%%%%%%%%%%%%%%%%%%%%%%%%%%%%%%%%%%%%%%%%

%%%%%%%%%%%%%%%%%%%%%%%%%%%%%% BIBLIOGRAPHY %%%%%%%%%%%%%%%%%%%%%%%%%%%%%%%%%%%%
%\input{bbl.tex}
%%%%%%%%%%%%%%%%%%%%%%%%%%%%%% END BIBLIOGRAPHY %%%%%%%%%%%%%%%%%%%%%%%%%%%%%%%%

\section{Acknowledgments} \label{sec:ack}
%%%%%%%%%%%%%%%%%%%%%%%%%%%%%% ACKNOWLEDGMENTS %%%%%%%%%%%%%%%%%%%%%%%%%%%%%%%%%
We would like to acknowledge helpful discussions with Dr. S. Litvinov, Dr. J. Zavadlav and Dr. E. Cruz-Chu. We would like to acknowledge the computational time at Swiss National Supercomputing Center (CSCS) under the project s659. We gratefully acknowledge support from the European Research Council (ERC) Advanced Investigator Award (No. 2-73985-14).
%%%%%%%%%%%%%%%%%%%%%%%%%%%%%% END ACKNOWLEDGMENTS %%%%%%%%%%%%%%%%%%%%%%%%%%%%%

\section{Author Contributions Statement} \label{sec:contrib}
L.K. ran the simulations, prepared the figures and tables, wrote the Results and Molecular Dynamics sections of the manuscript. G.A. prepared the single process HB code, the Supporting Information and the Bayesian Uncertainty Quantification text. P.A. prepared the LAMMPS script and guided the MD part of the research. P.K. wrote the Abstract, the Introduction and the Discussion sections. P.C. wrote the high performance computing implementation of the HB code and assisted in running the simulations. C.P. guided the Bayesian part of the research. All authors reviewed the manuscript.

\section{Additional Information}

\subsection{Competing financial interests}
The authors declare no competing financial interests.

\subsection{Availability of materials and data}
We use an open-source framework $\Pi$4U available at \verb|http://www.cse-lab.ethz.ch/software/Pi4U|.
The data we used comes from Ref.~\cite{Eisenstein:1942,Halpern:2010}.

\beginsupplement
\smaketitle

\section{Uncertainty quantification} \label{sec:uq_th_si}
This section provides a detailed description of the UQ theory used in the current work.

\subsection{Sampling the posterior distribution} \label{sec:sampling:posterior}
For the posterior distribution $\pr{\vp}[\vy, \M]$ which is known up to a normalizing constant $\pr{\vy}[\M]$, available Markov Chain Monte Carlo (MCMC) methods can be used to efficiently generate samples that quantify the uncertainty in $\vp$~\cite{Hastings:1970,Gilks:2005,Gamerman:2006,Geyer:1992,Peskun:1973}. In our work we use the Transitional Markov Chain Monte Carlo (TMCMC) algorithm~\cite{Ching:2007} with a slight modification on the MCMC proposal covariance used to overcome low acceptance rates in several runs of TMCMC which we observed. Instead of setting the covariance matrix to the scaled sample covariance of the previous generation, we use the landscape around the chain leaders to construct local covariance matrices. The sampling algorithm automatically tries to increase the radius of the neighborhood starting from 10\% of the domain size until the local covariance matrix is positive definite.

\subsection{Robust posterior prediction} \label{sec:prediction}
The uncertainty in the model parameters can be further propagated to the uncertainty in the QoI $\vec{y}$ produced by $f(\vx; \vp)$. Under the assumption of equation~\eqref{eq:likelihood} the probability of the model prediction conditioned on the parameters $\vp$ is given by $\pr{\vec{y}}[\vp, \M] = \N(\vec{y} \,|\, f(\vx; \vp), \C)$. The probability of the model prediction conditioned on the observations $\vy$ is known as the robust posterior prediction and is given by~\cite{Angelikopoulos:2012,Hadjidoukas:2015a}:
\begin{equation}
    \pr{\vec{y}}[\vy, \M] = \int \pr{\vec{y}, \vp}[\vy, \M] \, d\vp
    = \int \pr{\vec{y}}[\vp, \M] \, \pr{\vp}[\vy, \M] \, d\vp
    \approx \frac{1}{N_s} \sum_{k=1}^{N_s} \pr{\vec{y}}[\vp^{(k)}, \M] \cnt
\end{equation}
where $\vp^{(k)} \sim \pr{\vp}[\vy, \M]$ and $N_s$ is sufficiently large.

When a new QoI $\vec{z}$ produced by $g(\vx; \vp)$ is considered, $\C$ is unknown and only the parametric uncertainty is propagated into $g(\vx; \vp)$. Namely, one should estimate the density of $\sum_{k=1}^{N_s} g(\vx; \vp^{(k)})$, where $\vp^{(k)} \sim \pr{\vp}[\vy, \M]$ and $N_s$ is sufficiently large.

\subsection{Model selection} \label{sec:model:selection}
The Bayesian framework allows one to select the probabilistic model which best fits the data. The criterion for the model selection comes from the Bayes' theorem, which computes the probability of a probabilistic model $\M$ as
\begin{equation}
	\pr{\M}[\vy] = \frac{ \pr{\vy}[\M] \, \pr{\M} }{ \pr{\vy} } \cnt
\end{equation}
where $\pr{\M}$ is the prior PDF of the model $\M_i$ and the evidence $\pr{\vy}[\M]$ of model $\M$ is computed as a by-product of TMCMC.

\subsection{Hierarchical Bayesian models}
\begin{figure*}[hbtp]
    \centering
    \includegraphics[width=\textwidth]{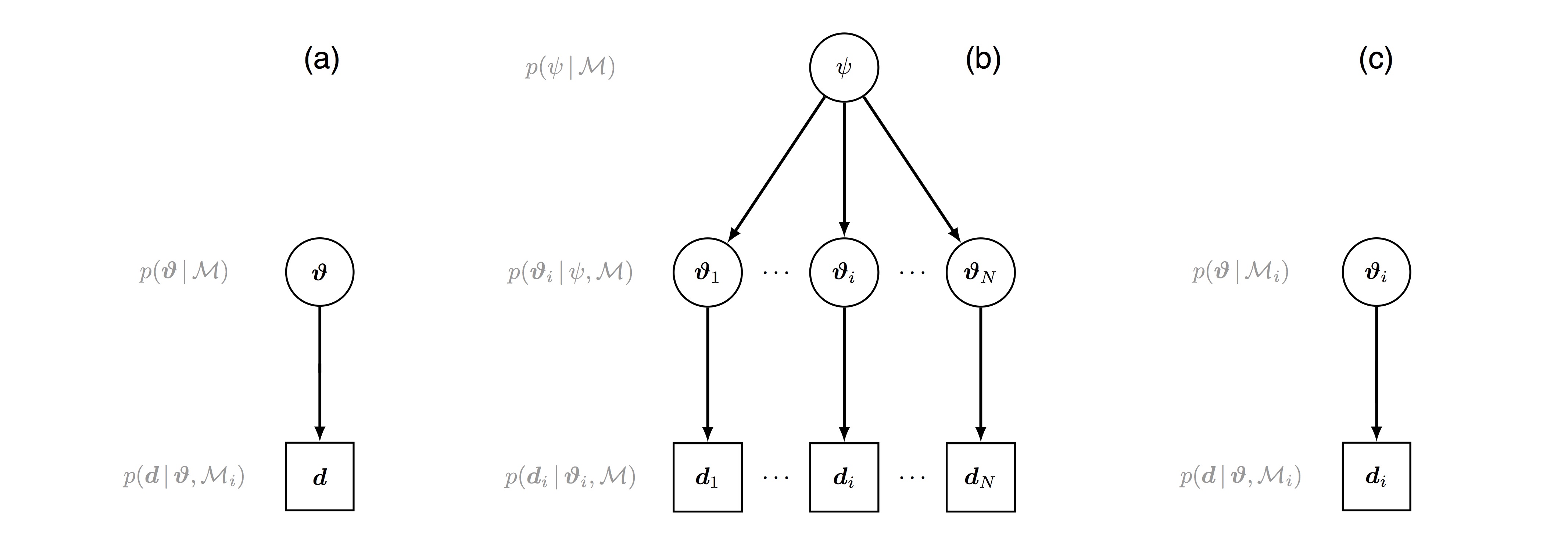}
    \caption{Bayesian networks: (a) simple network, (b) HB network, (c) network for the $i$-th dataset.}
    \label{fig:dag}
\end{figure*}

In our work we follow the methodology developed in \cite{Wu:2016}. We assume that data comes split in $N$ different datasets: $\vvy=\{\vy_1,\ldots, \vy_N\}$ and the likelihood in the probabilistic model $\M_i$ is $\N(\vy_i \,|\, f(\vx; \vp_i), \C_i)$. We assume that the probability of $\vp_i$ depends on a hyper-parameter $\vhp \in \R^{N_\hp}$ and is given by a PDF $\pr{\vp}[\vhp, \M]$, where $\M$ corresponds to the graph describing the relations between $\vhp$, $\vp_i$ and $\vy_i$, see Fig.~\ref{fig:dag}.
Our goal is to obtain samples from the posterior distribution, $\pr{\vp_i}[\vvy, \M]$, where $\vvy = \{\vy_1, \ldots, \vy_N\}$:
\begin{equation} \label{eq:h:lik:a}
    \pr{\vp_i}[\vvy, \M] = \int \pr{\vp_i}[\vhp, \vvy, \M] \, \pr{\vhp}[\vvy, \M] \, d\vhp \stp
\end{equation}
The dependency assumptions from Fig.~\ref{fig:dag} allow to simplify: $\pr{\vp_i}[\vhp, \vvy, \M] = \pr{\vp_i}[\vhp, \vy_i, \M]$, and equation~\eqref{eq:h:lik:a} can be rewritten using the Bayes' theorem:
\begin{equation} \label{eq:h:lik:b}
    \pr{\vp_i}[\vvy, \M] = 
    \int \frac{ \pr{\vy_i}[\vp_i, \vhp, \M] \, \pr{\vp_i}[\vhp, \M] }{ \pr{\vy_i}[\vhp, \M] } \, \pr{\vhp}[\vvy, \M] \, d\vhp \stp
\end{equation}
Since $\pr{\vy_i}[\vp_i, \vhp, \M] = \pr{\vy_i}[\vp_i, \M]$, equation~\eqref{eq:h:lik:b} simplifies to
\begin{equation}
    \pr{\vp_i}[\vvy, \M] = 
      \quad \pr{\vy_i}[\vp_i, \M] \int \frac{ \pr{\vp_i}[\vhp, \M] }{ \pr{\vy_i}[\vhp, \M] } \, \pr{\vhp}[\vvy, \M] \, d\vhp \stp
\end{equation}
Finally, the posterior distribution (\ref{eq:h:lik:a}) can be approximated as
\begin{equation}
    \pr{\vp_i}[\vvy, \M] \approx \frac{ \pr{\vy_i}[\vp_i, \M] }{ N_s } \sum_{k=1}^{N_s} \frac{ \pr{\vp_i}[\vhp^{(k)}, \M] }{ \pr{\vy_i}[\vhp^{(k)}, \M] } \cnt
\end{equation}
where, $\vhp^{(k)}\sim \pr{\vhp}[\vvy, \M]$ and $N_s$ is sufficiently large. Thus, in order to obtain $\vp_i$ samples, we first have to sample the probability distribution $\pr{\vhp}[\vvy, \M]$, which, according to Bayes' theorem, is equal to
\begin{equation}
    \pr{\vhp}[\vvy, \M] = \frac{ \pr{\vvy}[\vhp, \M] \, \pr{\vhp}[\M] }{ \pr{\vvy}[\M] } \cnt
\end{equation}
where $\pr{\vhp}[\M]$ is the prior PDF on $\vhp$ and $\pr{\vvy}[\M]$ is the normalizing constant. Exploiting the dependency assumption of Fig.~\ref{fig:hb} we see that
\begin{equation}
	\pr{\vvy}[\vhp, \M] = \prod_{i=1}^{N} \pr{\vy_i}[\vhp, \M] \cnt
\end{equation}
and the likelihood of $i$-th dataset can be expressed according to the total probability theorem as
\begin{equation} \label{eq:h:lik:c}
    \pr{\vy_i}[\vhp, \M] = \int \pr{\vy_i}[\vp_i, \M] \, \pr{\vp_i}[\vhp, \M] \, d\vp_i \stp
\end{equation}
Here we introduce the model $\M_i$ described in Fig.~\ref{fig:dag}. The posterior distribution of this model will be used as instrumental density for important sampling. Under the modeling assumption $\pr{\vy_i}[\vp_i, \M] = \pr{\vy_i}[\vp_i, \M_i]$ (see \cite{Wu:2016}) and the use of the Bayes' theorem, equation~\eqref{eq:h:lik:c} is written as
\begin{equation}
       \pr{\vy_i}[\vhp, \M] =
       \int \frac{ \pr{\vp_i}[\vy_i, \M_i] \, \pr{\vy_i}[\M_i] }{ \pr{\vp_i}[\M_i] } \, \pr{\vp_i}[\vhp, \M] \, d\vp_i \cnt
\end{equation}
or, equivalently, as
\begin{equation}
    \pr{\vy_i}[\vhp, \M] =
	\pr{\vy_i}[\M_i] \int \frac{ \pr{\vp_i}[\vhp, \M] }{ \pr{\vp_i}[\M_i] } \, \pr{\vp_i}[\vy_i, \M_i] \, d\vp_i \stp
\end{equation}
Finally, equation~\eqref{eq:h:lik:c} can be approximated as
\begin{equation}
    \pr{\vy_i}[\vhp, \M] \approx
	\frac{ \pr{\vy_i}[\M_i] }{ N_s } \sum_{k=1}^{N_s} \frac{ \pr{\vp_i^{(k)}}[\vhp, \M] }{ \pr{\vp_i^{(k)}}[\M_i] } \cnt
\end{equation}
where $\vp_i^{(k)} \sim \pr{\vp_i}[\vy_i, \M_i]$ and $N_s$ is sufficiently large. Note that in general $N_s$ can be different for each data set $\vy_i$. The advantage of this approach is that the likelihoods $\pr{\vy_i}[\vp_i, \M_i],\, i=1,\ldots,N$,  which are the most expensive part of the computations, are not re-evaluated for each $\vhp$.

\section{Information about hyper-parameter models for hierarchical inference} \label{sec:hb_info}
\paragraph{Inference for LJ 6-12} We assume
\begin{equation}
    \pr{\vp}[\vhp, \M] = \prod_{j=1}^3 \pr{\p_j}[\vhp, \M]
\end{equation}
and consider the following two models: 
\begin{itemize}
    \item[1)] uniform: $\pr{\p_j}[\vhp, \M] = \U( \p_j \,|\, \hp_{2j-1}, \hp_{2j-1} + \hp_{2j})$, where $\U(\xi | a, b)$ is the uniform distribution of $\xi$ with parameters $a, b$ and $\M$ is set to $\MU$,
    \item[2)] log-normal: $\pr{\p_j}[\vhp, \M] = \LN(\p_j \,|\, \hp_{2j-1}, \hp_{2j})$, where $\LN(\xi | a, b)$ is the log-normal distribution of $\xi$ with parameters $a, b$ and $\M$ is set to $\MLN$.
\end{itemize}
The prior distribution on the hyper-parameters is modeled as independent uniform,
\begin{equation} \label{eq:hp_12}
    \pr{\vhp}[\M] = \prod_{j=1}^6 \U( \hp_j \,|\, a_j^{\M}, b_j^{\M}) \cnt
\end{equation}
where $\M \in \{\MU, \MLN\}$ and the constants $a_j^{\M}, b_j^{\M}$ are given in Table~\ref{tab:hb_12}, along with the values of the log-evidences for the two models. The model $\U$ is according to the Bayesian model selection criterion, an order of magnitude more plausible and thus will be used for the further inference.
\begin{table*}[hbtp]
	\begin{minipage}{0.45\textwidth}
		\caption{$HB_{12, R}$ inference: lower bound and width for each of the hyper-parameters defined in equation~\eqref{eq:hp_12}, log-evidences for each hyper-prior model.}
		\label{tab:hb_12}
		\centering
		\begin{tabular}{l l l l}
			\hline \hline
			& $ \M=\MU$ & $\M=\MLN$ \\
			\hline
			$[a_{1}^{\M}, b_{1}^{\M}]$ & [0.0, 3.0]  & [-1.000, 2.0] \\
			$[a_{2}^{\M}, b_{2}^{\M}]$ & [0.0, 7.0]  & [ 0.001, 2.5] \\
			$[a_{3}^{\M}, b_{3}^{\M}]$ & [3.0, 3.4]  & [-3.000, 0.5] \\
			$[a_{4}^{\M}, b_{4}^{\M}]$ & [0.0, 2.0]  & [ 0.001, 4.0] \\
			$[a_{5}^{\M}, b_{5}^{\M}]$ & [0.0, 0.2]  & [-3.500, 0.5] \\
			$[a_{6}^{\M}, b_{6}^{\M}]$ & [0.0, 1.0]  & [ 0.001, 2.5] \\
			\hline
			Log-ev. & -19.1401 & -22.0198 \\
			\hline \hline
		\end{tabular}
	\end{minipage}
	\begin{minipage}{0.5\textwidth}
		\caption{$HB_{\pp, R}$ inference: lower bound and width for each of the hyper-parameters defined in equation~\eqref{eq:hp_p}, log-evidences for each hyper-prior model.}
		\label{tab:hb_free}
		\centering
		\begin{tabular}{l l l l}
			\hline \hline
			& $\M=\MU$ &  $\M=\MLN$ & $\M=\MTN$ \\
			\hline
			$[a_{1}^{\M}, b_{1}^{\M}]$ & [0.0, 3.00] & [-1.000, 2.0] & [0.0500, 10.0] \\
			$[a_{2}^{\M}, b_{2}^{\M}]$ & [0.0, 7.00] & [ 0.001, 2.5] & [0.0010, 3.30] \\
			$[a_{3}^{\M}, b_{3}^{\M}]$ & [3.0, 3.40] & [-3.000, 0.5] & [3.0000, 4.00] \\
			$[a_{4}^{\M}, b_{4}^{\M}]$ & [0.0, 2.00] & [ 0.001, 4.0] & [0.0010, 0.30] \\
			$[a_{5}^{\M}, b_{5}^{\M}]$ & [6.0, 7.00] & [-2.500, 1.5] & [6.0000, 12.0] \\
			$[a_{6}^{\M}, b_{6}^{\M}]$ & [0.0, 10.0] & [ 0.001, 2.5] & [0.0010, 2.00] \\
			$[a_{7}^{\M}, b_{7}^{\M}]$ & [0.0, 0.20] & [-3.500, 0.5] & [0.0001, 1.00] \\
			$[a_{8}^{\M}, b_{8}^{\M}]$ & [0.0, 1.00] & [ 0.001, 2.5] & [0.0010, 0.30] \\
			\hline
			Log-ev. & -24.0202 & -27.3889 & -25.3927 \\
			\hline \hline
		\end{tabular}
	\end{minipage}
\end{table*}
\paragraph{Inference for LJ 6-$\pp$} We assume
\begin{equation}
    \pr{\vp}[\vhp, \M] = \prod_{j=1}^4 \pr{\p_j}[\vhp, \M]
\end{equation}
and consider the following three models: 
\begin{itemize}
    \item[1)] uniform: $\pr{\p_j}[\vhp, \U] = \U( \p_j \,|\, \hp_{2j-1}, \hp_{2j-1} + \hp_{2j})$, where $\U(\xi | a, b)$  is the uniform distribution of $\xi$ with parameters $a, b$ and $\M$ is set to $\MU$,
    \item[2)] log-normal: $\pr{\p_j}[\vhp, \M] = \LN(\p_j \,|\, \hp_{2j-1}, \hp_{2j})$, where $\LN(\xi | a, b)$ is the log-normal distribution of $\xi$ with parameters $a, b$ and $\M$ is set to $\MLN$,
    \item[3)] truncated normal: $\pr{\p_j}[\vhp, \M] = \TN(\p_j \,|\, \hp_{2j-1}, \hp_{2j})$, where $\TN(\xi | a, b)$ is the truncated normal distribution of $\xi$ with parameters $a, b$ and $\M$ is set to $\MTN$.
\end{itemize}
The prior distribution on the hyper-parameters is modeled as independent uniform,
\begin{equation} \label{eq:hp_p}
    \pr{\vhp}[\M] = \prod_{j=1}^8  \pr{\hp_j}[\M]  =  \prod_{j=1}^8 \U( \hp_j \,|\, a_j^{\M}, b_j^{\M}) \cnt
\end{equation}
where $\M \in \{\MU, \MLN, \MTN\}$ and the constants $a_j^{\M}, b_j^{\M}$ are given in Table~\ref{tab:hb_free}, along with the values of the log-evidences for the three models. As it can be seen, the uniform model is the most plausible one.

\section{Posterior parameter distribution for LJ 6-12 and LJ 6-$\pp$} \label{sec:ind_post}
This section presents the posterior distributions for LJ 6-12 and LJ 6-$\pp$ obtained in HB TMCMC runs for each thermodynamic condition as well as distribution for the quantum dimer-based Bayesian inference. Each plot contains all the TMCMC samples of the last stage and is made as follows: histograms of marginal distributions of parameters are shown on the diagonal, projections of the samples to all possible 2-d subspaces in the parameter space colored by the log-likelihood values are given above the diagonal, the corresponding densities constructed via a bivariate kernel estimate are depicted below the diagonal. Green star shows the parameters from Ref.~\cite{Barker:1971}, green square indicates the parameters from Ref.~\cite{White:1999}, and green circle marks the parameters from Ref.~\cite{Rahman:1964,Rowley:1975}.

\end{document}